# Parameterized Hardware Architecture for Frame Synchronization at all Noise Levels

Dimitris Nikolaidis

*Abstract*— Frame synchronization is the act of discerning the first bit of a valid data frame inside an incoming transmission. This is particularly important in high-noise environments where the communication channel significantly alters transmitted signals. Sync word frame synchronization is a subcategory of synchronization methods where sync words are detected through digital correlation. Despite its simplicity, this method has been overlooked in literature in favor of more sophisticated and mathematically more optimal solutions. In this article we employ binary sync-word correlation-based synchronization to achieve near perfect frame synchronization at any noise level. The proposed architecture leverages XNOR gates, adder and comparator tree structures to detect sync words that are placed in front of the frames through digital correlation. The tree structures are circuit elements that mimic binary trees in form and provide the summation (adder tree) or the maximum/minimum (comparator tree) of a set of binary numbers as output. Due to their minimalistic nature, synchronization can be implemented practically for very large sync word sizes (>500 bit) with multigigabit bit rates (>20 Gbps) and very high accuracy ($10^{-5}$ synchronization error when the bit error rate on the bitstream is close to 0.3) on commercial FPGAs. The architecture also delivers the payload of the frames to its output as an extra function.

## I. INTRODUCTION

Frame synchronization is the ability of a receiving system to accurately detect incoming data frames. In general, this is achieved by detecting the position of the first bit of the frame and is very important in noisy environments where signals are significantly distorted. Excessive miss detection of the start of the frame leads to excessive frame loss and throughput deterioration.

One of the oldest frame synchronization schemes was introduced by Barker in 1956 [1] and utilizes digital correlation to detect the frames. Barker attached special sync words on the frames called Barker sequences that possess the optimal autocorrelation property. This means that the sidelobes of the autocorrelation operation are as small as possible, enabling more accurate detection. Massey in 1972 [2] expanded this idea by proving mathematically that the optimal metric to discern the position of the frame is not just the correlation value alone but the correlation value minus an energy term. This came to be known as the maximum likelihood rule and has been a staple in frame synchronization.

Since then, as communication technology progressed, so did frame synchronization schemes. There were altered maximum likelihood rule versions centered around examining multiple frames [3] or sequentially examining a certain time interval for the start of the frame [4]. There were also schemes targeted at variable length frame synchronization [5]. Of particular importance are the schemes which perform incoherent frame synchronization [6] where there is frequency uncertainty [7] [8] at the receiver.

In recent years, frame synchronization has expanded even more in newly established areas. There have been frame synchronization solutions where machine learning has been employed [9] [10] and solutions aimed at specific modulation types such as LoRa [11] which is primarily used for IoT applications [12]. New maximum likelihood metrics were also employed to provide synchronization in FSO links [13]. Moreover, with the advancement in hardware technology, error decoding has become efficient enough to allow for joint frame synchronization and error correction. Error codes such as Polar [14] and LDPC [15] are used to perform simultaneous synchronization and correction.

Even though the range of schemes in literature provide solutions for a plethora of applications, most of them are highly specialized, involve complex mathematics and do not offer a clear path to practical implementation. The digital architecture presented in this article aims to provide an easy to understand and implement universal solution that offers near perfect synchronization in almost all noise levels. This is achieved with the least possible amount of hardware resources for the largest possible bit rates. As a result, the architecture can offer bit rates up to 40 Gbps while consuming less than 50% of the resources of a medium-scale commercially available FPGA.

The proposed architecture accepts the received demodulated bit stream in segments of $q$ parallel bits. It utilizes XNOR gates, adder and comparator trees to perform digital correlation on very long sync words and detects them by monitoring the correlation peaks. Sync words are of size $n$ (multiple of $q$) and are attached to every frame. The demodulated bitstream enters the architecture and is correlated with the predefined expected sync word. Correlation is achieved by using XNOR gates on the stream and sync word and by adding the outputs of the XNOR gates. Addition is executed by $q$ parallel adder trees which are basic circuit elements that consist of adders (binary) and their forms resemble binary trees. In each level of the tree, numbers are added by two until the summation of the entire set is calculated. A comparator tree is used to discern the maximum out of all $q$ correlation values in the bitstream. It follows the form of the adder tree though instead of adders it has

Dimitris Nikolaidis is with the National Technical University of Athens email: (dimnikolaidis@mail.ntua.gr)

comparators. The detected maximum is compared with a threshold and serves as an indication that a sync word is detected. It is then used to capture the corresponding frame data. The correlation operation between two binary vectors can be seen on table 1.

TABLE I
DIGITAL CORRELATION BETWEEN TWO 6-BIT VECTORS. THE OUTCOME OF THE SUMMATION INDICATES HOW MANY BITS ARE THE SAME.

| Vector A | 1 | 1 | 0 | 1 | 0 | 1 |
|---|---|---|---|---|---|---|
| Vector B | 1 | 0 | 0 | 0 | 0 | 1 |
| XNOR | 1 | 0 | 1 | 0 | 1 | 1 |
| Sum | 100 (4) | | | | | |

The pure correlation method is a suboptimal method that presents worse accuracy results in theory compared to other methods; however, by building the architecture with simple elements we can increase the size of the sync word greatly without restrictive hardware consumption. This allows us to essentially use brute force and achieve synchronization with error in the range of $10^{-5}$ under bit error rate on the bitstream close to 0.3(one in three bits are transmitted wrong). This level of accuracy has not been presented by any other method or implementation. While the architecture targets coherent frame synchronization (symbol synchronization has been achieved) it will have become apparent by the end that the accuracy of the method is so high that it does not matter whether receiver and transmitter are perfectly synchronized.

The architecture presented in this article is the updated generalized version of the architecture presented in [16]. In the old version the size of the sync word $n$ must be the same as the size of the parallel bit input $q$, $n=q$. This is an important disadvantage when one needs high accuracy and large sync words but does not require a high bit rate. The current version removes this restriction and enables the user to choose input size $q$ independently of $n$, optimizing accuracy, hardware and power consumption. It is however required that $q$ is smaller than or equal to $n$, $n \geq q$. A more advanced FPGA platform is also used to implement more hardware demanding architecture versions with better accuracy and bit rate. In fact, it offers synchronization error rates close to $10^{-5}$ even when the bit error rate on the bit stream is in the range of 0.3-0.25.

## II. PRELIMINARIES

To simplify the presentation of the method and architecture some assumptions have been made. Firstly, the architecture accepts $q$ parallel bits in every clock cycle from the received demodulated bitstream as input. This means that instead of receiving one bit at frequency $F$ it receives $q$ bits at frequency $F/q$. This parallelization allows the architecture to handle bit rates ($F$) much higher than what the maximum clock of the platform allows (ASIC/FPGA). Secondly, the sync words are transmitted in front of the frame data and are of size $n$ bits where $n$ is a multiple of $q$. In the absence of frames, the channel transmits only 0s (idle state). The length of the idle state varies and does not affect the architecture. The sync words are random binary sequences of bits which are created prior to communication. They are random in the sense that they are created randomly (equal chance of 0 or 1 to appear) but once they are generated, they are known to both receiver and transmitter. Predefined sync words are also given as input to the architecture. The visual representation of the input stream divided into sections of $q$ bits including sync words, frame data and the idle state can be seen in Figure 1. Time flows from right to left so that in a block of $q$ bits (bits $q−1$ down to 0) bit 0 is transmitted earlier than bit $q−1$.

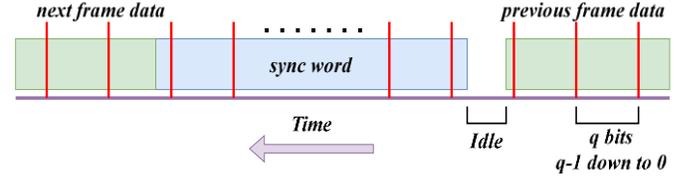

**Fig. 1.** The flow of time on the demodulated input bitstream. The data rate is $F$ and $q$ bits enter the architecture every clock cycle with frequency $F/q$.

Thirdly, for the remainder of the paper, the index of an individual bit on a set of possible placements in the input bitstream will be referred to as position, and the index of the first bit of a group of bits will be referred to as location i.e. location of the sync word means the index of the first bit of the sync word.

## III. ARCHITECTURE

The architecture consists of three modules. Isolation Window, Parallel Correlation and Fata Capture and can be seen in Figure 2. The Isolation Window Module isolates a certain area of the incoming demodulated bitstream and funnels it to the Parallel Correlation Module for processing. It only contains D flip-flops that constitute a larger register of size $n+2×q$ bits ($n$ size of the sync word and $q$ size of the parallel input) which is larger than the minimum needed to identify the sync word, but it is necessary to facilitate both detection and capture of the frame data.

The Parallel Correlation Module correlates the input from the Isolation Window Module with the predetermined sync word and outputs the most likely location of the sync word. It achieves this by using XNOR gates on the respective bits and then adding the outputs of the gates with pipelined parallel adder trees depending on the size of parallelization. The outputs of the adder trees are then driven through a pipelined comparator tree that determines which out of all the possible locations has the highest correlation value and is the most likely location for the sync word. The module also features a delay register which synchronizes outputs with respective inputs to enable frame capture. Its outputs are the most likely location of the sync word $m$ along with its respective correlation value *Sum* and the synchronized delayed input.

Frame Capture is not traditionally seen in other frame synchronization schemes, but it was implemented here as an extra feature. The term capture refers to providing frame data as output of the architecture. The module uses the most likely location of the sync word and the contents of the delay register sent by Parallel Correlation to accurately deduce the position of the first bit of the valid data of the frame. It compares the correlation value (*Sum*) to a given threshold and when the threshold is surpassed it assumes that the sync word is detected. Frame data is funneled directly from the delay register to the output of the module at a rate of $q$ bits per clock cycle, which is the data rate of the input. The module also provides a *Valid Data* signal which signifies capture is underway.

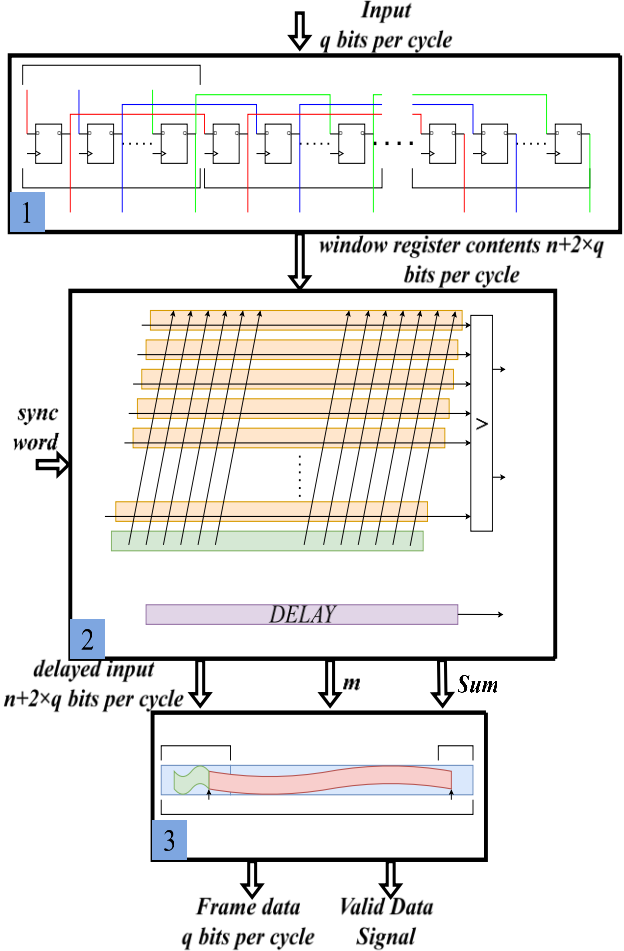

**Fig. 2.** The frame synchronization architecture. Module 1 is Isolation Window, Module 2 is Parallel Correlation and Module 3 is Frame Capture. The architecture accepts $q$ parallel bits every cycle and outputs $q$ frame data bits per cycle (assuming a frame was detected).

*A. Isolation Window Module*

The Isolation Window Module accepts the input bitstream in segments of $q$ bits per clock cycle. Since the location of the sync word can be any out of these $q$ bits and considering that the size of the sync word itself is $n$ bits, the window module register must be at least of size $q+n-1$ where the possible locations for the sync word are $n-1$ down to 0, $n$ down to 1, $n+1$ down to 2, $n+2$ down to 3… $n+q-2$ down to $q-1$. The window module register consists of $n/q+2$ slots of $q$ bits for a total of $n+2\times q$ bits. This size is bigger than the minimum required, however it is necessary to both detect the sync word and capture the frame data without other modifications. Detection of the sync word happens only on the first $q$ bits ($q-1$ down to 0) of the register. Input enters the last slot of the register (slot $n/q+1$) and in every cycle the contents of one slot move one slot to the right until they reach the first slot, slot 0. This implies that positions that are $q$ bits apart are connected serially. Figure 3 presents the circuit.

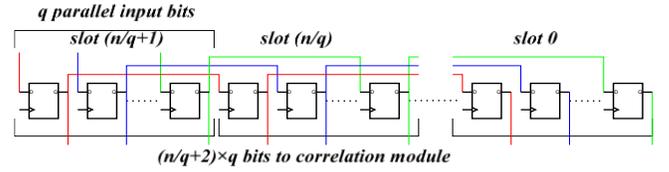

**Fig. 3.** Circuit of the Isolation Window Module register. It has $n/q+2$ slots of $q$ bits for a total of $n+2\times q$ bits. Color coded wires indicate that registers are serially connected.

The contents of the register are funneled to the Parallel Correlation Module to determine the most likely location of the sync word.

*B. Parallel Correlation Module*

The Parallel Correlation Module accepts the contents of the Isolation Window Module register and correlates the predefined sync word with the sets of bits for every possible location ($n+q-2$ down to $q-1$, $n+q-3$ down to $q-2$…..$n$ down to 1, $n-1$ down to 0). At any time $q$ correlation operations are executed in parallel. The set of bits for a possible location (for example $n+q-3$ down to $q-2$) are driven through the XNOR gates together with the respective bits of the sync word. The outputs of the gates are then added with a pipelined adder tree (registers are placed between the levels). In total $q$ adder trees and $n\times q$ XNOR gates are needed to perform correlation for all possible locations. The outputs of the adder trees enter a pipelined comparator tree that determines the value and position of the maximum correlation value that is the most likely location of the sync word. The comparator tree follows the binary tree form of the adder tree but instead of adders it has comparators. In every stage half of the values advance to the next level and at the end the highest value of the set appears as the output of the entire comparator tree. The index of each value (in respect to the set) is kept in a parallel register and advances together with its corresponding value. It appears at the end as a second output of the tree. In the design, the *Sum* signal represents the value of correlation and $m$ represents the index which coincides with the location of the sync word. Both the adder trees and the comparator tree are pipelined with registers in between levels and their combined latency is ceil($\log_2 n$)+ceil($\log_2 q$) clock cycles. The correlation circuit is presented in Figure 4.

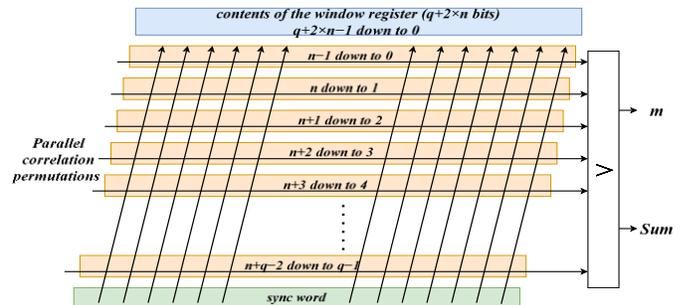

**Fig. 4.** The architecture of the Parallel Correlation Module. The diagonal lines represent the XNOR operation of a sync word bit with the bit in the respective position of each parallel permutation. The horizontal lines represent the adder tree summation of the outputs of the XNOR gates for every permutation. The ">" is the comparator tree which selects the maximum value *Sum* along with its index $m$. $m$ is also the index of the first bit of the sync word inside the window register.

Together with the correlation circuit the Module features a delay register to synchronize the two outputs *Sum* and *m* with the Isolation Window register contents that produce them. It has ceil($\log_2 n$)+ceil($\log_2 q$) (latency) slots of $n+2\times q$ (size of isolation window register) bits each. The bits in the first slot (slot 0) mirror the contents the register before ceil($\log_2 n$)+ceil($\log_2 q$) clock cycles which means that they are the corresponding inputs of outputs *Sum* and *m* of the same clock cycle. The delay register together with *Sum* and *m* are used by the Frame Capture Module to deliver the frame data to the output of the architecture.

*B. Frame Capture Module*

Frame Capture Module is the last module in the chain and the simplest of the three. It delivers the frame data to the output of the architecture and controls the *Valid Data* signal that signifies capture. Detection of the frame occurs by monitoring the maximum correlation value (*Sum*) of window register. When it surpasses the given threshold at clock cycle *t* the module assumes that the frame has been found and begins the capture process. If the size of the frame is a multiple of *q* ($k\times q$), bits in positions $m+n+q-1$ down to $m+n$ are captured for the next *k* cycles (*q* bits per clock cycle). The *Valid Data* signal is activated during the capture process. Notice that since $m \leq q-1$ we have $m+n+q-1 \leq n+2\times q-2$, so in order to both detect and capture frame data the delay register needs at least $n+2\times q-1$ bit positions. Since *n* is a multiplier of *q*, $n+2\times q$ ($n+2\times q-1$ down to 0) is the closest multiplier of *q* to $n+2\times q-1$ and also the size of the window register. Figure 5 presents the position of the first bit of data inside slot 0 of the delay register.

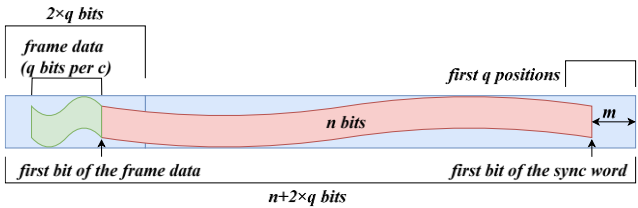

**Fig. 5**. The position of the first bit of the sync word (red) and frame data(green) within the window register when the start of the sync word is detected on the first *q* (*q*−1 down to 0) bits of slot 0.

Since the capture module has only one chance to capture the frame data, the threshold needs to be appropriately set to provide as much accuracy as possible. There is the possibility to keep monitoring the correlation value for some clock cycles after the threshold is surpassed and if a higher correlation value is found to restart the capture process from the beginning, however this method is very hard to implement practically and does not offer much benefit when the sync words are very large. For sync words that are above 300 bits in length, the correlation at the correct position is always much higher than other positions so it is easily distinguished.

IV. IMPLEMENTATION AND RESULTS

The architecture was implemented for 9 pairs of *n*, *q* values. Table 2 presents the results of implementation in terms of LUTs, FFs, maximum operating frequency, power and calculated throughput ($q \times FPGA\_operating\_frequency$). The platform of implementation was the xilinx board KCU 116 and the tool used was VIVADO 2023.

For all pairs except (*n*=780, *q*=78), (1020,68) and (1020,85) the maximum operating frequency is 450MHz. For (780,78), (1020,68) and (1020,85) which correspond to the largest circuits, the achieved operating frequency is 400 MHz. The high operating frequencies achieved can be attributed to the minimalist nature and pipelined form of the architecture that consists only of XNOR gates, simple adder structures and comparators. The tool implemented the adders on the fabric and no DSP units were utilized. Utilization percentages for LUTs range from 19.19% to 84.50% and for Flip-Flops (FFs) range from 13.74% to 60.38%. LUTRAMS utilization % is lower than 1% for all implementations. Power on chip was found to be between 4.54 and 17.647W. The frequency reduction on the three largest circuits can probably be attributed to sub optimized mapping of the large sized circuit. Notice that the version that supports the highest bit rate (540,90) does not consume most resources. It is however less accurate due to the shorter sync word length. The relation between the size of parallelization *q* and throughput is obvious ($q \times FPGA\_operating\_frequency$) and explains why versions of similar *q* have similar throughput regardless of *n*. Both *n* and *q* affect hardware consumption, *n* being the size of the one adder tree unit and parallelization *q* the number of adder trees used in the correlation module.

TABLE II
IMPLEMENTATION RESULTS

| n, q | LUTs (%) | FFs (%) | LUTRAMS (%) | Max. Op. F. (MHz) | Throughput (Gbps) | Power (W) | P/T (W/Gbps) |
|---|---|---|---|---|---|---|---|
| 540,36 | 41637 (19.19) | 59628(13.74) | 301(0.30) | 450 | 16.2 | 4.544 | 0.280 |
| 540,60 | 69345(31.96) | 98963(22.80) | 480(0.48) | 450 | 27 | 7.232 | 0.267 |
| 540,90 | 104354(48.09) | 148197(34.15) | 720(0.72) | 450 | 40.5 | 10.687 | 0.263 |
| 780,30 | 49672(22.89) | 71590(16.49) | 210(0.21) | 450 | 13.5 | 5.214 | 0.386 |
| 780,52 | 86229(39.74) | 123488(28.45) | 364(0.36) | 450 | 23.5 | 8.878 | 0.377 |
| 780,78 | 123422(56.88) | 184839(42.59) | 546(0.54) | 400 | 31.2 | 10.528 | 0.337 |
| 1020,34 | 73279(33.77) | 105446(24.30) | 80(0.08) | 450 | 15.3 | 7.470 | 0.488 |
| 1020,68 | 146688(67.61) | 209812(48.35) | 149(0.15) | 400 | 27.2 | 14.695 | 0.540 |
| 1020,85 | 183343(84.50) | 262035(60.38) | 170(0.17) | 400 | 34 | 17.647 | 0.519 |

Considering the P/T (Power/Throughput) results for all versions, P/T is influenced more by the size of the sync word $n$ as versions of the same $n$ have similar P/T regardless of parallelization $q$. This is expected and can be shown if we consider the following. Since the FPGA maximum operating frequency $F/q$ is set any change to the bit rate $F$ means proportional change to the degree of parallelization $q$. Moreover, parallelization $q$ represents the number of adder trees used in the design, so it also proportionally affects power. As a result, changes to the bit rate $F$ (or throughput) eventually lead to proportional changes to power leading to similar P/T ratio for designs where $n$ is the same.

The testing of the architecture was facilitated through emulation. Three iterations of the architecture (540,60), (780,52), (1020,68) with different sync word sizes were tested since the accuracy of synchronization is only determined by the size of the sync word. Synthetic data was prepared in MATLAB and given to the circuit as input in VIVADO post implementation simulation. The data consisted of 245098 frames with attached sync words of appropriate size (540,780,1020) for each architecture version. The data were randomly generated by MATLAB functions, modulated with 16QAM and driven through the AWGN channel for all integer SNR dB values within the range of $[-8,-2]$. The capture module threshold that is used to detect the correlation peak was set to 0.65 of the size of the sync word (351, 507, 663 respectively). The size of each frame was set to $300 \times q$ ($k$=300) which is above the average IP packet (12000 bits) for all versions (18000, 15600, 20400). The results of frame synchronization error rate i.e. the rate at which frames are lost can be seen in Figure 6 for every SNR value for all architecture versions. BER (bit error rate) of the bit stream was also added to put into perspective how accurate the correlation with long sync words is.

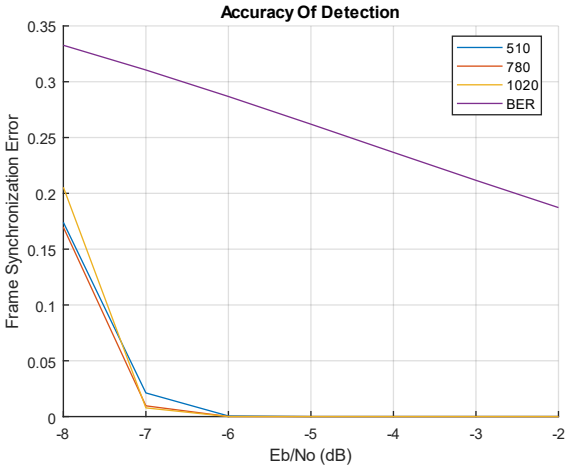

**Fig. 6.** Accuracy of detection for each architecture version. The BER curve signifies how accurate all versions are relative to the number of errors on the bitstream.

Synchronization errors (miss-detections) appeared only for SNR values -8dB, -7dB, -6dB. The measured accuracy values of the three versions at SNR values -8dB, -7dB, -6dB are 0.174016, 0.021187, 0.000538 for 510, 0.169663, 0.0096313, $6.12 \times 10^{-5}$ for 780 and 0.205440, 0.007784, $1.63 \times 10^{-5}$ for 1020 respectively. From -5dB and after no errors were detected which means that the chance of losing a frame is less than $\frac{1}{245098}=4.08 \times 10^{-6}$ (four frames in a million are lost) and negligible. This is a significant result since at this noise level for 16QAM, the bitstream presents a bit error rate (BER) equal to 0.26, much higher than what other methods can process. As expected, the length of the sync word affects the synchronization accuracy. (1020,68) performs better than both (780,52) and (540,60) at $-7$dB and $-6$dB and (780,52) performs better than (540,60). At -8dB (1020,68) performs worse than the others probably due to the extreme percentage of errors (0.33) which render the size of the sync word irrelevant to accuracy.

In relation to other frame synchronization methods, the architecture provides many magnitudes better accuracy for higher bit rate. [17] and [18] are comprehensive publications which offer comparisons between the accuracy of joint frame synchronization with LDPC codes and Masey's modified max likelihood correlation metric. The best result presented is about $10^{-2}$ around 0 dB which is far below our reported results of $1.63 \times 10^{-5}$ accuracy at $-6$ dB. Publications regarding other methods report similar accuracy results (<100) [19][3][11]. In terms of hardware implementation, max likelihood and correlation methods do not offer any architecture in literature. The only method that presents frame synchronization circuits is the LDPC joint error correction-frame synchronization method in the form of LDPC decoders. According to survey [20] these circuits have lower bit rates while being algorithmically more complex [21]. More specifically the highest reported bit rate for an LDPC decoder on an FPGA is reported to be 698 Mbps with 66,885 LUTs [22] while the simplest proposed architecture version achieves 16.2 Gbps with 41,637LUTs.

In our tests each frame was attached to a single sync word, yet the architecture can also be used in other ways. One may send multiple sync words to synchronize the bitstream and then start sending data frames. The architecture provides both the value and position of the correlation peak so following this method can be done without modifications. Determining the best method depends on the characteristic of the channel, the resources available and constraints and will be explored in future work. It is also important to note that since the architecture accepts the sync word as an input and the sync word itself is randomly generated (equal chance of 0 and 1) the architecture can also be used as a pseudo-cryptography mechanism to conceal information being transmitted. Without knowledge of the preamble, it is impossible to detect the frame.

## IV. CONCLUSION

In this paper we have introduced a novel frame synchronization architecture based on very long randomly pre-generated sync word correlation. The architecture leverages only XNOR gates, adder trees and comparators to provide synchronization with few resources. It correlates the incoming demodulated bit stream with the expected sync word and deduces the location of the frame data on the stream using the correlation values detected. The architecture functions at high operational frequencies and together with long sync word correlation it provides extremely accurate frame synchronization at multigigabit bit rates (13.5, 27.2, 40.5). It performs above all other schemes proposed in terms of accuracy and bit rate. It can also conceal transmitted frames due to the randomness of the sync word acting as a pseudo cryptography mechanism. The board Xilinx kcu116 was used as the platform of implementation with a wide range of power and hardware consumption affected by the size of the sync word, parallelization and frequency.